\documentclass[aps,twocolumn,amsfonts,amsmath,amssymb,showkeys,nobibnotes,nofootinbib,floatfix,a4paper]{revtex4-1}

\usepackage{graphicx} 
\usepackage{amsfonts} 
\usepackage{amsmath} 
\usepackage{amssymb} 
\usepackage{hyperref} 
\usepackage{siunitx} 
\usepackage{booktabs} 
\usepackage{cleveref} 
\usepackage{color} 
\usepackage{ulem}

\providecommand{\e}[1]{\ensuremath{\times 10^{#1}}}

\begin{document} 
	
	\title{The multi-layer network nature of systemic risk and its implications for the costs of financial crises} 

	\author{Sebastian Poledna$^1$}
	\author{Jos\'{e} Luis Molina-Borboa$^4$}
	\author{Seraf\'{i}n Mart\'{i}nez-Jaramillo$^4$}
	\author{Marco van der Leij$^{5,6,7}$}
	\author{Stefan Thurner$^{1,2,3}$}
	\email{stefan.thurner@meduniwien.ac.at}

	\affiliation{
	$^1$Section for Science of Complex Systems, Medical University of Vienna, Spitalgasse 23, A-1090, Austria\\
	$^2$Santa Fe Institute, 1399 Hyde Park Road, Santa Fe, NM 87501, USA\\
	$^3$IIASA, Schlossplatz 1, A-2361 Laxenburg, Austria\\
	$^4$Direcci\'{o}n General de Estabilidad Financiera, Banco de M\'{e}xico, Ave. 5 de Mayo 2, Ciudad de M\'{e}xico, Distrito Federal, M\'{e}xico\\
	$^5$CeNDEF, University of Amsterdam, Valckeniersstraat 65-67, 1018 XE Amsterdam, The Netherlands\\
	$^6$Research Department, De Nederlandsche Bank, Westeinde 1, 1017 ZN Amsterdam, The Netherlands\\
	$^7$Tinbergen Institute, Gustav Mahlerplein 117, 1082 MS Amsterdam, The Netherlands}

	\begin{abstract}
		The inability to see and quantify systemic financial risk comes at an immense social cost. Systemic risk in the financial system arises to a large extent as a consequence of the interconnectedness of its institutions, which are linked through networks of different types of financial contracts, such as credit, derivatives, foreign exchange and securities. The interplay of the various exposure networks can be represented as layers in a financial multi-layer network. In this work we quantify the daily contributions to systemic risk from four layers of the Mexican banking system from 2007-2013. We show that focusing on a single layer underestimates the total systemic risk by up to 90\%. By assigning systemic risk levels to individual banks we study the systemic risk profile of the Mexican banking system on all market layers. This profile can be used to quantify systemic risk on a national level in terms of nation-wide expected systemic losses. We show that market-based systemic risk indicators systematically underestimate expected systemic losses. We find that expected systemic losses are up to a factor four higher now than before the financial crisis of 2007-2008. We find that systemic risk contributions of individual transactions can be up to a factor of thousand higher than the corresponding credit risk, which creates huge risks for the public. We find an intriguing non-linear effect whereby the sum of systemic risk of all layers underestimates the total risk. The method presented here is the first objective data driven quantification of systemic risk on national scales that reveal its true levels. 
	\end{abstract}


	\maketitle 

\section{Introduction} \label{intro}
Systemic risk (SR) in financial markets is the risk that a significant fraction of the financial system can no longer perform its function as a credit provider and collapses. In a more narrow sense, SR is a notion of contagion or impact that starts from the failure of a financial institution (or a group of institutions) and propagates through the financial system, and potentially to the  real economy \citep{De-Bandt:2000aa,BIS:2010aa}. Systemic risk in financial markets generally emerges through two mechanisms, either through the synchronization of behavior of agents (fire sales, margin calls, herding), or through the interconnectedness of agents. The former can be measured by a potential capital shortfall during periods of synchronized behavior, where many institutions are simultaneously distressed \citep{Adrian:2011aa,Acharya:2010aa,Brownlees:2012aa,Huang:2012aa}. The latter is a consequence of the network nature of financial claims and liabilities \citep{Eisenberg:2001aa,Boss:2004aa}. Network-based SR is potentially extremely harmful because of the possibility of cascading failure, meaning that the default of a financial agent may trigger defaults of others. Secondary defaults might cause avalanches of defaults percolating throughout the entire network and can potentially wipe out the financial system by a de-leveraging cascade \citep{Minsky:1992aa,Fostel:2008aa,Geanakoplos:2010aa,Adrian:2008aa,Brunnermeier:2009aa,Thurner:2012aa,Caccioli:2012aa,Poledna:2014ab,Aymanns:2014aa}. The fear of cascading failure is generally believed to be the reason why institutions under distress are often bailed out at tremendous public costs \citep{Klimek:2014aa}. On the regulators' side, in response to the financial crisis of 2007-2008, broader attention is now directed to SR. A consensus for the need for new financial regulation -- including a potential re-design of the financial world -- is emerging \citep{Aikman:2013aa}. In the currently discussed regulation framework of Basel III the importance of networks is recognized \citep{BIS:2010aa,Georg:2011aa}.

These developments have spurred research on SR and financial networks. It has been shown that the topology of financial networks can be associated with probabilities for systemic collapse \citep{Haldane:2011aa,Roukny:2013aa}. In particular, network centrality measures have been identified as appropriate measures to quantify SR by various groups \citep{Boss:2004aa,Puhr:2012aa,Markose:2012aa,Caballero:2012aa,Billio:2012aa,Minoiu:2013aa,Thurner:2013aa}. A disadvantage of centrality measures is that the SR value for a particular node has no clear interpretation as a measure for expected losses in the case of a cascading failure event. A variant of a centrality measure that solves this problem is the so-called {\em DebtRank}, which is a recursive method to quantify the systemic relevance of financial nodes in terms of losses \citet{Battiston:2012aa}. This improvement, achieved by the DebtRank, has inspired recent work on financial SR, involving real data \citep{Poledna:2014aa} and agent based models \citep{Thurner:2013aa}.

Despite the tremendous importance of SR and the research efforts devoted to the topic, so far there are no reliable quantitative indices that quantify SR on a \emph{national} and \emph{temporal} basis. Indices that have sometimes been used to estimate SR in markets -- such as volatility indices (such as VIX), or spreads of credit default swaps (such as CDX) -- are poor proxies because they are clearly incapable of taking cascading defaults into account. As a consequence these proxies greatly underestimate the true levels of SR in economies.  

In this work we develop a number of potentially practical methods to quantify SR in financial multi-layer networks. First, we extend the notion of systemic importance in financial network to multi-layer networks. This makes it possible to assess SR contributions from various layers of financial networks. Second, we develop a risk measure to quantify the expected loss due to SR, that takes cascading into account by explicit use of the financial network topologies on a daily scale. This risk measure extends the notion of systemic importance to a national level and allows us to compare  SR levels of economies over time, and to identify trends and historical events. In this sense the measure can be used as an indicator or a SR index. In particular it becomes possible to compare SR levels and their related potential costs, before and after the recent crisis. Third, building on the work of \citet{Poledna:2014aa}, we use the risk measure to quantify the marginal contribution of individual exposures in financial networks to the overall SR. This allows us to extend the notion of systemic importance from financial institutions to individual exposures. In particular it allows us to quantify the expected loss due to SR associated with every individual exposure of financial institutions.

This work is based on a unique data set containing various types of daily exposures between the mayor Mexican financial intermediaries (banks) over the period 2004-2013 (for this work we use data from 2007-2013). Data is collected and owned by the Banco de M\'{e}xico and has been extensively studied under various aspects \citep{Martinez-Jaramillo:2010aa,Lopez-Castanon:2012aa,Martinez-Jaramillo:2014aa}. Here we focus on banks that interact simultaneously in four different markets, generating four different types of exposures: (unsecured) interbank credit, securities, foreign exchange and derivative markets. Hence, institutions are connected by contracts of four different types. Different contract types can be seen as distinct network layers. A collection of various networks linking the same set of nodes is called a multi-layer or \emph{multiplex network}. The interplay of the various exposure networks can be represented as layers in a financial multi-layer network. The data further contains the capitalization of banks for every month. With this data we quantify the SR contributions of the individual layers and estimate the mutual influence of one layer of exposures on the others. 

We obtain a series of practically relevant results. First, we show that focusing on a single exposure layer individually underestimates the total SR by up to 90\%. When focusing on all the layers, we find an intriguing non-linear effect that the sum of SR from all layers underestimates the total SR. Second, we show that market-based SR indicators systematically underestimate expected systemic losses. Third, we find that current expected systemic losses are up to a factor four higher now than they were before the financial crisis of 2007-2008. Fourth, we find that SR contributions of individual transactions can be up to a hundred times higher than the corresponding credit risk, which creates huge risks for the public. 

The method presented here is the first objective, data driven quantification of SR on national scales that reveals its true levels on a temporal basis. 

\section{Related literature}
Our work contributes to existing literature on SR and financial multi-layer networks. In recent years, several contributions to the statistical understanding of multi-layer networks and their dynamics have appeared in a broad and general context \citep{Szell:2010aa,Nicosia:2013aa,Kim:2013aa}. Especially the use of network similarity measures, node- and link correlations, and link-overlap measures have turned out to be useful tools to identify and quantify interactions between layers \citep{Nicosia:2013aa,Kim:2013aa,Szell:2010aa}. The various layers of a financial multi-layer network consist of credit (borrowing-lending relationships, consisting of counterparty exposures and implicit relationships, such as roll-over of overnight loans), insurance (derivative) contracts, collateral obligations, market impact of overlapping asset portfolios and the network of cross-holdings (holding of securities or stocks of other banks). Research on financial networks has mainly focused on a single layer; mostly, on direct lending networks between financial institutions \citep{Upper:2002aa,Boss:2004aa,Boss:2005aa,Soramaki:2007aa,Iori:2008aa,Cajueiro:2009aa,Bech:2010aa,Fricke:2014aa,Iori:2014aa}, but also on the network of derivative exposures \citep{Markose:2012aa,Markose:2012ab}, and on the network of common asset exposures \citep{Greenwood:2014aa}. 

Research on financial \emph{multi-layer} networks has only appeared recently. \citet{Leon:2014aa} study the interactions of financial institutions on different financial markets in Colombia. \citet{Bargigli:2013aa} study the interaction in the Italian interbank market between financial network layers of short- and long-term bilateral lending, both secured and unsecured. \citet{Bargigli:2013aa} and \citet{Leon:2014aa} are however not directly concerned with measuring SR.

\citet{Bluhm:2014aa} consider an agent-based model of a multi-layer interbank network, incorporating different contagion channels -- that is, from common asset exposure, direct lending exposures and fire sales. They are not concerned with the interaction between the individual layers. On the other hand, \citet{Montagna:2013aa} do consider the contribution of individual contagion layers to SR. Their agent-based model consists of three layers: long-term direct lending exposures, short-term direct lending exposures, and common asset exposures. Calibrating the model on end-of-2011 balance sheet data from 50 European banks, they obtain a similar result to us, namely that SR measured on the combined multi-layer network can be much larger than the aggregate SR from the individual network layers. However, \citet{Montagna:2013aa} do not have actual bilateral network data at their disposal; instead, they consider the multi-layer network structure as a free parameter in their calibration exercises, and SR depends on the network structure considered. By using daily data on the exact bilateral exposures between banks in the Mexican financial network, our work, however, does not allow for this freedom. The use of high frequency longitudinal data allows a detailed analysis of the fluctuations of different SR contributions over time. Furthermore, \citet{Montagna:2013aa} do not consider exposures from derivatives, cross-holdings of securities and FX transactions, which in the Mexican data set turn out to be the dominant exposures, much more relevant than short-term or long-term unsecured deposits and loans. Moreover, whereas \citet{Montagna:2013aa} consider as systemic risk measure the number of failing banks in a default cascade, we consider a systemic risk measure based on DebtRank, which allows for a simple interpretation of SR in terms of expected losses, and a simple comparison to market-based systemic risk measures. Thus, the availability of more granular and extensive data allows us to take the analysis of \citet{Montagna:2013aa} a considerable step further.  

In this context several risk measures for SR have been proposed that focus (mainly) on statistics of losses, accompanied by a potential shortfall during periods of synchronized behavior, where many institutions are simultaneously distressed \citep{Adrian:2011aa,Acharya:2010aa,Brownlees:2012aa,Huang:2012aa}. In particular, four statistical measures have been proposed recently: conditional value-at-risk (CoVaR), systemic expected shortfall (SES), systemic risk indices (SRISK) and distressed insurance premium (DIP). CoVaR is defined as the value at risk (VaR) of the financial system, conditional on institutions being in distress. The contribution to SR of an institution is the difference between CoVaR, conditional on the institution being in distress, and CoVaR in its median state \citep{Adrian:2011aa}. SES measures the propensity to be undercapitalized, given that  the system as a whole is undercapitalized \citep{Acharya:2010aa}. SES is related to leverage and the marginal expected shortfall (MES). SRISK is closely related to SES and as such, a function of the size of an institution, its degree of leverage, and its MES \citep{Brownlees:2012aa}. DIP measures the price of insurance against systemic financial distress in the banking system and is closely related to the expected shortfall \citep{Huang:2012aa}. None of these measures take cascading defaults into account. 

\section{Quantification of Systemic Risk in multi-layer networks}
\subsection{The financial multi-layer network --  different exposure types} 
Banks interact in different markets and generate different types of exposures. Banks issue securities that are later bought by other banks. By holding these securities, banks expose themselves to other banks. Foreign exchange transactions can lead to large exposures between banks. Their exposures are associated with settlement risk. Another market activity that can lead to considerable exposures is trading in financial derivatives.

We analyze four different types ($\alpha=1,2,3,4$) of financial exposures: `derivatives', `securities', `foreign exchange' and `deposits \& loans'. In \cref{data} we explain in detail how the exposure types are obtained from the Mexican data set. We use the following notation for different exposure types: The size of every exposure of type $\alpha$ of institution $i$ to institution $j$ at time $t$ is given by the matrix element $L_{ij}^{\alpha}(t)$. $\alpha=1,2,3,4$ labels the layers `derivatives', `securities', `foreign exchange' and `deposits \& loans', respectively. We use the convention to write liabilities in the rows (second index) of matrix $L$, so that the entries $L_{ij}^{\alpha}(t)$ at a given day $t$ are the liabilities bank $i$ has towards bank $j$. If the matrix is read column-wise (transpose of $L$) we get the assets or exposures of banks. Although the layers considered in this paper arise from different types of financial risk, the links between nodes in all the layers have the same meaning as the total loss that might arise for an institution, as the consequence of the default of another. The concept and dimension (dollars) of exposure is the same for all links in all layers: it is the total loss that one institution would suffer if a given counterparty defaulted (in any given layer), for details see \cref{data}. 

\subsection{DebtRank -- quantification of systemic risk at the institutional level} \label{multidr}
DebtRank was originally suggested as a recursive method to determine the systemic relevance of nodes within financial networks \citep{Battiston:2012aa}. It is a quantity that measures the fraction of the total economic value $V$ in the network that is potentially affected by the distress of an individual node $i$, or by a set of nodes $S$. For details see \cref{debtrank_section}. The DebtRank of a set of nodes $S$ that is initially in distress is denoted by $R_{S}$. In those cases where only one node $i$ is initially under distress (the set $S$ contains only one node $i$) we denote the Debt Rank of that node by  $R_i$. 

DebtRank values can be computed for each layer of a multi-layer network, $L_{ij}^{\alpha}$ separately. For DebtRanks of layer $\alpha$, $R_i^{\alpha}$, $L_{ij}$ in \cref{impact} is simply replaced by $L_{ij}^{\alpha}$. The economic value at each layer, that is necessary for the  computation of $R_i^{\alpha}$, is given by $v_{i}^{\alpha}=L_{i}^{\alpha}/\sum_{j}L_{j}^{\alpha}$, where $L_{i}^{\alpha}=\sum_{j}L_{ji}^{\alpha}$. For  a multi-layer network, DebtRank can be calculated from the combined liability network $L_{ij}^{\rm comb} = \sum_{\alpha} L_{ij}^{\alpha}$. We refer to the DebtRank of the combined liability network as $R_i^{\rm comb}$ and the total economic value $V^{\rm comb}=\sum_{i}L_{i}^{\rm comb}$ is given by total interbank assets in all layers combined, see \cref{ecovalue1}. 

To allow a comparison of $R_i^{\alpha}$ between different layers, $R_i^{\alpha}$ must be shown as a percentage of the total economic value $V^{\rm comb}$ of interbank assets in all layers combined (\cref{ecovalue1}). The normalized DebtRank for layer $\alpha$ is therefore defined as 
\begin{equation}
	\label{debtrank_layer} \hat R_i^{\alpha} = \frac{V^{\alpha}}{V^{\rm comb}} R_i^{\alpha} \quad, 
\end{equation}
where $V^{\alpha}=\sum_{i}L_{i}^{\alpha}$ is the total economic value of the interbank assets in the layer $\alpha$.

\subsection{Quantification of systemic risk  at the country level} \label{sr_profile_sec} 
We define the {\em SR-profile} of a country at time $t$ as the rank-ordered normalized DebtRank values $\hat R_i^{\alpha}$ for all financial institutions in a country. The SR-profile shows the distribution of systemic impact across institutions throughout a country. The institution with the highest SR level is to the very left. We denote the number of institutions in a country by $b$. It is natural to define the {\rm average DebtRank} as a quantity that captures the SR of the entire economy (with $b$ institutions)  at a given time, 
\begin{equation}
	\bar R ^{\alpha}(t) = \frac{1}{b} \sum_{i=1}^{b} \hat R_i^{\alpha}(t) \quad . 
\end{equation}
For the combined network, $\hat R_i^{\alpha}$ is replaced by $R_i^{\rm comb}$, and we write $\bar R ^{\rm comb}(t)$ for the combined average DebtRank. Note that $\bar R ^{\alpha}(t)$ depends on the network topology of the various layers (or the combined network) only, and is independent of default probabilities, recovery rates, or other variables. 

The precise meaning of the DebtRank as the fraction of the total economic value in a network allows us to define the {\em expected systemic loss} for the entire economy, which is the size of the loss, multiplied by the probability of that loss occurring \citep{Poledna:2014aa}. To compute the expected systemic loss, we first consider the simple case where only one institution $i$ can default and all other $b-1$ institutions survive.  In this case the expected loss is given by ${\rm EL}_{i}^{\rm syst}({\rm one\,\, default}) = V \cdot p_i \cdot (1-p_1) \cdot \ldots \cdot (1-p_{i-1}) \cdot (1-p_{i+1}) \cdot \ldots \cdot (1-p_{b}) \cdot R_i$, where $p_i$ is the probability of default of institution $i$, and $(1-p_j)$ the survival probability of $j$. The general case occurs when we also consider possible joint defaults, meaning that a set of institutions $S$ go into distress. Taking into account {\em all} possible combinations of defaulting and surviving institutions, we arrive at a combinatorial expression of the expected loss for an economy of $b$ institutions  
\begin{equation}
	{\rm EL}^{\rm syst} = V \sum_{S \in \mathcal{P}(B)} \prod_{i \in S} p_i \prod_{j \in B \setminus S} (1-p_j) \: R_S \quad, \label{totEL_normalized} 
\end{equation}
where $\mathcal{P}(B)$ is the power set of the set of financial institutions $B$, and $R_S$ is the DebtRank of the set $S$ of nodes initially in distress. $R_{\emptyset}$, the DebtRank of the empty set is defined as zero. The reason being, that by definition of DebtRank, $R_S \leq 1$, the value obtained in \cref{totEL_normalized} cannot exceed the total economic value. Note that the same arguments apply equally for individual network levels or the combined multi-layer network, where ${\rm EL}^{\rm syst}$ can be calculated with $R_i^{\alpha}$ or $R_i^{\rm comb}$, in combination with the corresponding  economic values \cref{ecovalue1}. 

It is immediately clear that  \cref{totEL_normalized} is computationally feasible only for situations with relative small numbers of financial institutions. Computing the power set and calculating DebtRanks for all possible combinations of large financial networks is unfeasible. In \cref{el_approx} we derive a practical approximation for \cref{totEL_normalized}, 
\begin{equation}
	{\rm EL}^{\rm syst} \approx V \sum_{i=1}^{b} p_i \, R_i \quad. \label{totEL} 
\end{equation}
This approximation is certainly valid if the default probabilities are low ($p_i \ll 1$), or the interconnectedness is low ($R_i \approx v_i$). To show that the approximation is extremely precise as far as realistic scenarios are concerned, we compare the exact result of \cref{totEL_normalized} with the approximation \cref{totEL} for an economy of $15$ banks in \cref{el_approx}. Note that in principle \cref{totEL} can become larger than $V$. However, in \cref{el_approx} we show that the approximation is a maximum of $3.5\%$ off the exact value of \cref{totEL_normalized}, which is never larger than $V$. For larger sets of banks, such as the Mexican data set, we would run into computational difficulties in computing the 17 billion odd combinations. For the remainder we will, therefore, use \cref{totEL}. 

\subsection{Quantification of systemic risk of individual exposures}
We estimate the impact of individual daily exposures on SR. In particular we compare the credit risk (expected loss) of a single exposure of a given size to its impact on SR. The expected loss (credit risk) of bank $i$ is
\begin{equation}
	\label{expected_loss} {\rm EL}_{i}^{\rm credit}(t) = \sum_{j=1}^{b} p_j \, {\rm LGD}_j \, L_{ji} (t)\quad, 
\end{equation}
with $p_j$ being the default probability as above, ${\rm LGD}_j$ the loss-given-default of $j$, and $L_{ji}$ the exposure at default of $i$ to $j$. The marginal contribution of an individual exposure, $X_{kl}$ (matrix with precisely one non-zero element in line $k$ and row $l$, quantifying the exposure between banks $k$ and $l$) on credit risk, is the increase of credit risk of the bank with the additional exposure (risk taken by lender), 
\begin{equation}
	\Delta {\rm EL}^{\rm credit}=\sum_i\left[{\rm EL}_{i}^{\rm credit}( L_{ij}+X_{kl}) - {\rm EL}_{i}^{\rm credit}(L_{ij})\right] \quad. 
\end{equation}
Here ${\rm EL}_{i}^{\rm credit}( . )$ means that ${\rm EL}_{i}^{\rm credit}$ is computed from the network in the argument. 

The impact of individual daily exposures on SR (marginal contribution), $\Delta {\rm EL}^{\rm syst}$ has been defined in \citet{Poledna:2014aa}. The marginal contribution of an individual exposure, $X_{kl}$ (matrix with precisely one nonzero element for the exposure between $k$ and $l$) on ${\rm EL}^{\rm syst}$ is the difference of total expected systemic loss, 
\begin{multline}
	\Delta {\rm EL}^{\rm syst} = \sum_{i=1}^{b} p_i \left[ V(L_{ij}+X_{kl})  \,
	R_i(L_{ij}+X_{kl},C_i) \,   \right.  \\ \left. - V(L_{ij}) \, R_i(L_{ij},C_i) \right] \, \quad, \label{marginal_effect_simple} 
\end{multline}
where $R_i(L_{ij}+X_{kl},C_i)$ is the DebtRank and $V(L_{ij}+X_{kl})$ the total economic value of the liability network without the specific exposure $X_{kl}$. Clearly, a positive $\Delta {\rm EL}^{\rm syst}$ means that $X_{kl}$ increases total SR. In general, this risk is borne by the public. If the increase in SR and credit risk of individual transactions are equal, $\Delta {\rm EL}^{\rm syst}=\Delta {\rm EL}^{\rm credit}$, a default of the exposure would only affect one of the involved parties (lender), and would not involve any third party. For transactions where $\Delta {\rm EL}^{\rm syst}>\Delta {\rm EL}^{\rm credit}$ third parties will also be affected by the default. The deviation  of $\Delta {\rm EL}^{\rm syst}-\Delta {\rm EL}^{\rm credit}>0$ is a simple and clear indicator for the existence of an incentive problem, where costs to third parties are generated by bilateral exposures.

\section{Data} \label{data} 
The data used for this work is derived from a database on exposures at the Mexican Central Bank, built and operated with the specific purpose of studying contagion and SR. This project is maintained by the statistics unit at the financial stability general directorate at this institution. The statistics unit under the financial stability general directorate at Banco de Mexico gathers information and cross-validates it by using  daily, weekly and monthly regulatory reports, which are used for regulatory and supervisory purposes. An illustrative and important example is the case of the daily regulatory reports known as `operaciones de captaci\'on e interbancarias en moneda nacional y udis' (OCIMN), and `operaciones de captaci\'on e interbancarias en moneda extranjera' (OCIME). These reports contain every single funding transaction on a daily basis in local and foreign currency, which are used to compute the daily funding costs for each bank. From these two  regulatory reports it is possible to compute the exact daily unsecured exposures between banks, as well as more broadly, those between financial institutions like investment banks, brokerage houses, mutual funds and pension funds. In \citet{Solorzano-Margain:2013aa}, a stress testing study was carried out using an (extended set) of these exposures. Given the confidential nature of these transactions,  data is kept under strict access control and can only be used for regulatory, supervisory and financial stability purposes.

The present work is based on transaction data that is converted to bilateral exposures. The four exposure types are obtained in the following way. 

\subsection{Deposits \& loans}
Daily exposures arise from interbank deposits and loans in local and foreign currency, and from credit lines extended for settlement purposes. In the case of deposits and loans, the calculation of exposures is straightforward. Maturities and funding risk are not relevant in the context of this paper because we are only concerned with the quantification of the loss-given-default of a counterparty. Effects of funding risk and how it propagates among institutions can be found in \citet{Lee:2013aa}. The current exposures $L_{ij}^{\alpha=4}(t)$ are calculated by adding up all deposits and loans between bank $i$ and $j$. As is the case with most studies, we calculate the gross exposure instead of net exposure. 

\subsection{Securities cross-holdings}
Daily exposures also arise  from  cross-holding of securities between banks, securities lending, securities that are used as collateral, and from securities trading. Cross-holding of securities between banks means that bank $j$ is holding securities issued by bank $i$. We again use the gross exposure, because security contracts must be honored, even when the counterparty defaults. The daily cross-holdings gross exposures $L_{ij}^{\alpha=2}(t)$ are calculated by adding up all securities cross-holdings that exist between bank $i$ and $j$. 

\subsection{Derivatives}
Daily exposures arise from the valuation of derivatives transactions, including swaps, forwards, options and repo transactions. For the derivatives layer, for each type of derivative contract (swaps, forwards or options) between any two given banks, the contract is valuated and the resulting net exposure (at the contract level) is then calculated and assigned to the corresponding bank. Banks provide the information needed to perform a vanilla Black Scholes valuation of the derivative contracts. However, on the intra-month scale, banks themselves valuate most of the derivative contracts and Banco de M\'{e}xico verifies the valuation methodologies at each banks' risk offices in order to guarantee proper modeling. In contrast to securities, exposures from derivatives are netted by each type of derivative contract. There are detailed international agreements on the netting procedure in the case of failure of a counterparty. This means, for instance, that options with the same underlying security are added up on each side and the exposures is then assigned to the counterparty with positive net position. This process is replicated for each type of derivative with the same underlying security. The resulting net exposures are then added up to calculate the final exposure $L_{ij}^{\alpha=1}(t)$, arising from derivative contracts between bank $i$ and bank $j$. In contrast to other, more developed financial systems, derivatives in Mexico do not generate size-able exposures and no exotic derivatives are traded. Sophisticated derivative strategies are only defined and executed by the parent banks of the Mexican subsidiaries. 

\subsection{Foreign exchange}
As far as foreign exchange (FX) transactions are concerned, exposures reflect settlement risk (or Herstatt risk) -- the risk that a counterparty will not pay as obligated at the time of settlement. Mexican banks that are subsidiaries of internationally active banks are members of CLS (Continuous Linked Settlement), and are in the position to settle their FX transactions in a secured way. However, not all active banks in Mexico are in this situation and large exposures related to FX transactions do arise. If banks settle FX transactions between themselves by using the clearance service provided by CLS -- which eliminates time differences in settlement -- there is no exposure. Otherwise the exposure $L_{ij}^{\alpha=3}(t)$ includes both foreign currency receivable and foreign currency payable between bank $i$ and bank $j$. 

Finally, various balance sheet data on the 43 Mexican banks is also available, such as the capitalization measured at a monthly scale. 

\section{Results} \label{results} 
\subsection{The financial multi-layer network -- the Mexican banking system}
\begin{figure}
	\centering 
	\includegraphics[width=.49\textwidth]{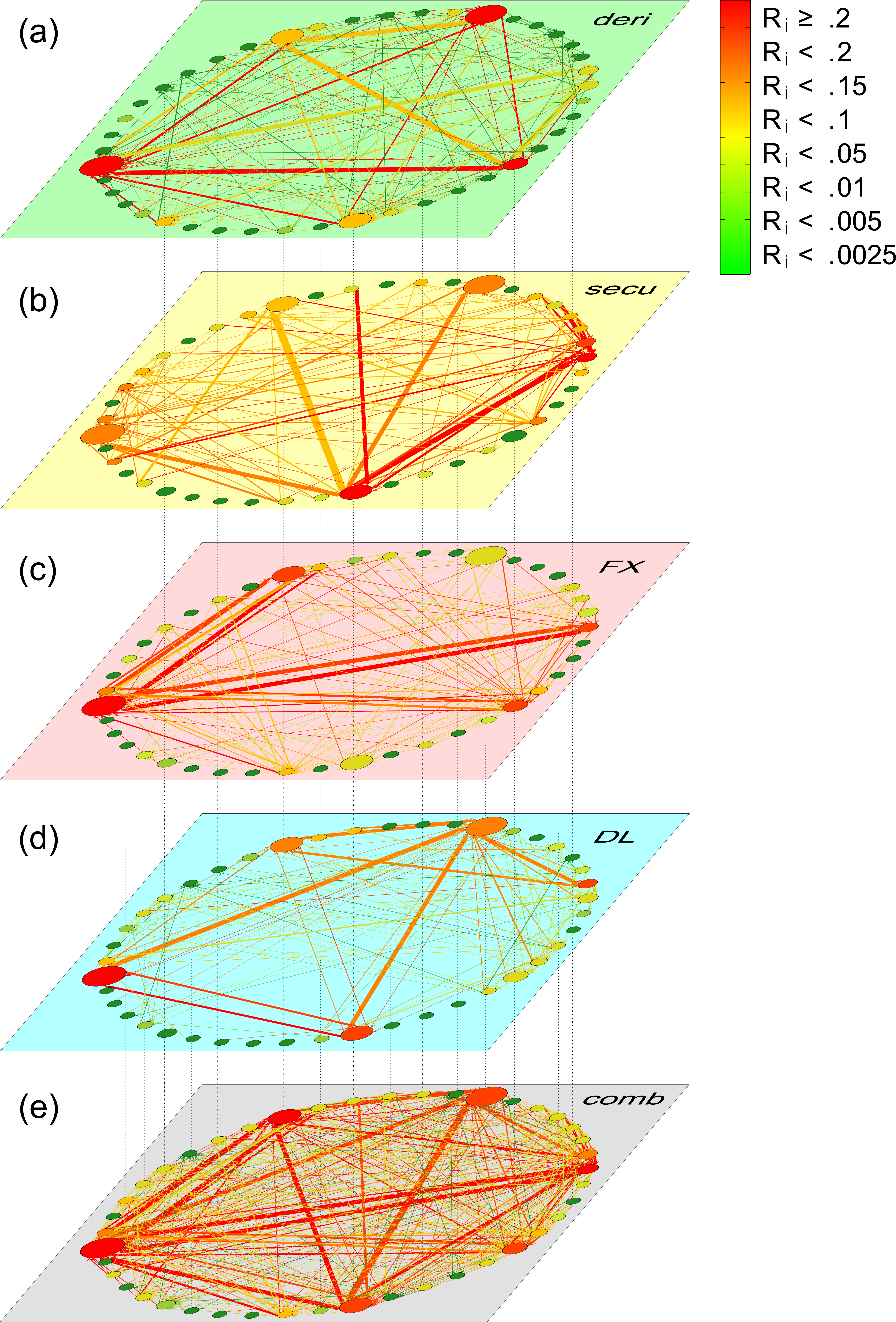} 
	\caption[Banking multi-layer network of Mexico on Sept 30 2013.]{Banking multi-layer network of Mexico on Sept 30 2013. (a) network of exposures from derivatives, (b) securities cross-holdings, (c) foreign exchange exposures, (d) deposits and loans and (e) combined banking network $L_{ij}^{\rm comb}(t)$. Nodes (banks) are colored according to their systemic impact $R_i^{\alpha}$ in the respective layer (see \cref{multidr}): from systemically important banks (red) to systemically safe (green). Node-size represents banks' total assets. Link-width is the exposure size between banks, link-color is taken from the counterparty.} \label{bnetwork} 
\end{figure}
\Cref{bnetwork} shows the various exposure layers of the Mexican banking network at Sept 30 2013. The derivative exposure network  is seen in the top layer (green), the second layer shows the exposures from securities cross-holdings (yellow), the third shows foreign exchange exposures (red). The fourth layer represents the interbank deposits and loans market (blue). Nodes are shown at the same position in all layers. Node-size represents the size of banks' total assets. Nodes $i$ are colored according to their systemic impact, as measured by the DebtRank, $R_i^{\alpha}$, in the respective layer (see \cref{multidr}). Systemically important banks are red, unimportant ones green. The width of links represents the size of the exposures in the layer; link-color is the same as the counterparty's  node color (DebtRank). The total exposure in layers $\alpha=2,3,4$, $\sum_{i,j} L_{ij}^{\alpha}(t) \approx 5\e{10}$ Mex\$, is similar in size. The total exposure of derivatives ($\alpha=1$) is smaller, $\sum_{i,j} L_{ij}^{1}(t) \approx 1\e{10}$ Mex\$. However, the number of links is larger in this layer. Note that the data for derivative exposures also contains exposures from so-called repo transactions; the respective amounts are  small (less than 2 \%) because the repo involves collateral. In \cref{bnetwork}(e) the combined exposures $L_{ij}^{\rm comb}(t) = \sum_{\alpha=1}^4 L_{ij}^{\alpha}(t)$ are shown. Classical network statistics for the multi-layer network are collected and discussed briefly in \cref{nw_statistics_sec}.

\begin{figure}
	\centering 
	\includegraphics[width=.49\textwidth]{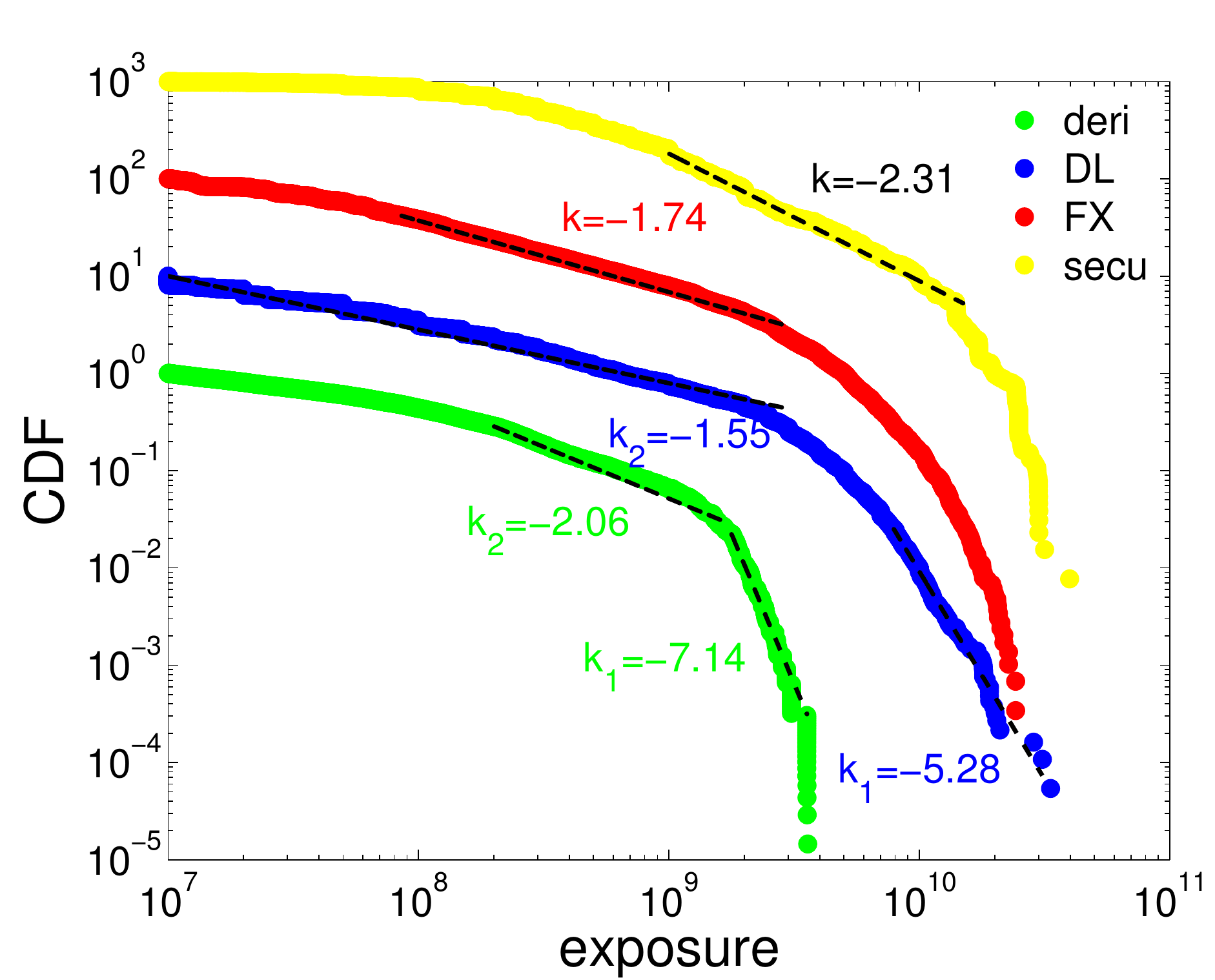}
	\caption{Cumulative distribution function (CDF) of the exposure sizes (in Mex\$) from the different layers, deposits and loans (DL), foreign exchange exposures (FX), derivatives (deri) and securities cross-holdings (secu). Data is aggregated from all days for the entire time span Jan 2 2007 to May 30 2013. Distributions are shifted vertically to avoid overlapping, by  factors of 10, 100, and 1000, respectively. Clearly, the distributions are not power-laws, however for comparison with previous literature, we report power laws fits (slopes) in several selected regions of the distributions.}
	\label{distr_fit} 
\end{figure}

The cumulative distribution function (CDF) of exposure sizes for the different layers $L_{ij}^{\alpha}(t)$ is presented in \cref{distr_fit}. Distributions are obtained by taking all exposures for every trading day in the observation period. Exposures from derivative holdings (green) are generally lower across the entire timespan. Deposits and loans (blue) are more frequent in small sizes; foreign exchange exposures (red) are typically the largest positions. The distribution of exposure sizes of securities cross-holdings (yellow) shows a higher variability for larger sizes compared to other layers. For clarity of the figure, we shifted the distributions for the deposits and loans, foreign exchange exposures,  and securities cross-holdings by the factors 10, 100, and 1000. Clearly, the observed distributions are neither power laws nor exponential functions. To formally determine whether the distributions expose power law tails, we conducted a standard goodness-of-fit test, which generates a $p$-value to quantify the plausibility of the hypothesis. We use an approach that combines a maximum-likelihood fitting method with a goodness-of-fit test, based on the Kolmogorov-Smirnov statistic and the likelihood ratios developed in \citet{Clauset:2009aa}. The $p$-values can be interpreted in the usual way. If $p<0.05$, then the null hypothesis is rejected at the 5\%-significance level. In that case there is sufficient statistical evidence that the distribution does not follow a power law for the identified region. The appropriate tests show the following results for each layer: Securities ($p<0.001$), FX ($p<0.001$), deposits \& loans ($p\approx0.0290$) and derivatives ($p<0.001$). All $p$-values are below the 5\%-significance level. In addition, for all exposure types, we use the goodness-of-fit based method described in \citep{Clauset:2009aa} to estimate the regions which can be fitted by a power. For deposits \& loans and derivatives we find two regions, which can be fitted by a power. The fitted values for deposits \& loans in the identified regions are $k_1\approx -5.28$ and $k_2\approx -1.55$ and for derivatives $k_1\approx -7.14$ and $k_2\approx -2.06$. For the securities layer and the FX layer we fitted only for one region, the values are $k\approx -2.31$ and  $k\approx -1.74$ respectively. 

\begin{table} 
	\begin{center}
		\begin{tabular}{l l l l l}
		 & $J_{\alpha\beta}$ & $\rho_{\alpha,\beta}^{\rm exp}$ & $\rho_{\alpha,\beta}^{\rm liab}$ & $\rho_{\alpha,\beta}^{R}$ \\ \hline
		DL:Deri    	& $0.096$ & $0.32^*$ & $0.20$ & $0.52$ \\
		DL:Secu   & $0.081$ & $0.40^*$ & $0.39$ & $0.63$ \\
		DL:FX  	& $0.082$ & $0.59^*$ & $0.16$ & $0.61$ \\
		Deri:Secu 	& $0.082$ & $0.04^*$ & $0.73^*$ & $0.19$ \\
		Deri:FX 	& $0.190^*$ & $0.56$ & $0.85^*$ & $0.63$ \\
		Secu:FX 	& $0.094$ & $-0.05^*$ & $0.66^*$ & $0.25$ 
		\end{tabular}
	\end{center}	
	\caption{Values for link-overlap (Jaccard coefficient $J_{\alpha\beta}$), and correlations of exposures (defined as $\sum_i L_{ij}^{\alpha}$) $\rho_{\alpha,\beta}^{\rm exp}$, liabilities ($\sum_j L_{ij}^{\alpha}$) $\rho_{\alpha,\beta}^{\rm liab}$ and DebtRank $\rho_{\alpha,\beta}^{R}$, between all possible combinations of two layers $\alpha$ and $\beta$, at Sep 30 2013. For the correlation of DebtRanks, $R_i^{\alpha}$ is calculated for the respective layers (see \cref{multidr}). Significant coefficients are marked with a star.} \label{corr_table}
\end{table}
To address the question of how similar the various exposures layers are, we compute the so-called link-overlap by calculating the Jaccard coefficient $J_{\alpha\beta}$ (see \cref{jaccard_section}) between two different layers $\alpha$ and $\beta$ for all possible pairs of layers. Furthermore, we compute the correlation coefficients between exposures (weighted in-degrees, $k_j= \sum_i L_{ij}^{\alpha}$), liabilities (weighted out-degrees, $k_i=\sum_j L_{ij}^{\alpha}$) and SR (DebtRank $R_i^{\alpha}$) for all banks, between all pairs of layers (see \cref{multidr}). Results are collected in \cref{corr_table}. Correlation coefficients and the Jaccard coefficient are computed for the multi-layer network observed at one representative trading day, i.e. at Sep 30 2013. In \cref{nw_statistics_sec} we show the evolution of pair-wise degree correlations from 2007 until 2013. The link-overlap (blue) between all pairs is relatively small. To test the significance of the observed link-overlap, we compare it to a null-model. For the randomized null-model we preserved the interbank assets or exposures of each bank (weighted in-degrees of $L_{ij}^{\alpha}$) and rewired the exposure to a random bank, similar to the method used by \citet{Maslov:2002aa}. Rewiring the interbank exposure to a random counterparty means that the liabilities (weighted out-degrees of $L_{ij}^{\alpha}$) are not preserved. Therefore total assets and equity capital of each bank remain unchanged. Note that it is not possible to preserve (weighted) in- and out-degree at the same time. We find that the link-overlap practically coincides with the null-model, meaning that if banks have business relations in one market it is not more likely that they also interact in other markets. Only the link-overlap between derivatives and foreign exchange show  slightly higher levels than the null-model, indicating that if two banks have exposure in securities the probability to have one in FX is marginally higher than in the null-model. 

High correlation coefficients $\rho_{\alpha,\beta}^{\rm exp}$ (see \cref{corr_section}) between total exposures of banks indicate that banks that have high (low) exposure in layer $\alpha$ have also high (low) exposure in layer $\beta$. Correlation coefficients $\rho_{\alpha,\beta}^{\rm exp}$ close to zero mean that total exposures of banks are not correlated in layer $\alpha$ and $\beta$. The correlations for liabilities of banks $\rho_{\alpha,\beta}^{\rm liab}$ and $\rho_{\alpha,\beta}^{\rm R}$ are interpreted in the same way. We find in \cref{corr_table} that $\rho_{\alpha,\beta}^{\rm exp}$ is close to zero for the pair (derivatives : securities), meaning that they are almost uncorrelated. We find negative correlations for the layer (securities : FX), meaning that exposures that are high in securities imply small exposures in FX and vice versa. Correlations for liabilities $\rho_{\alpha,\beta}^{\rm liab}$ are high for the pairs (derivatives : securities), (derivatives : FX) and (securities : FX), and low for (DL : derivatives) and (DL : FX). Compared to the null-model, correlation coefficients $\rho_{\alpha,\beta}^{\rm exp}$ for all pairs are significant, with the single exception of the pair (derivatives : FX). Here $\rho_{\alpha,\beta}^{\rm exp}$ coincides with the null-model. Correlations for liabilities $\rho_{\alpha,\beta}^{\rm liab}$ are significant for the pairs (derivatives : securities), (derivatives : FX) and (securities : FX). Finally, the correlation of SR at the bank level is $\rho_{\alpha,\beta}^{\rm R} > 0.5$ for all pairs except for (derivatives : securities) and (securities : FX), where correlations are  small. These latter pairs are different in the sense that their total exposures and systemic impact are practically uncorrelated. All the others layers are strongly correlated. Note that we cannot compare correlation results of SR at the bank level with the null-model because it is impossible to preserve (weighted) in- and out-degrees at the same time.

\subsection{Quantification of Systemic Risk in the Mexican banking system}
\begin{figure}
	\centering 
	\includegraphics[width=.49\textwidth]{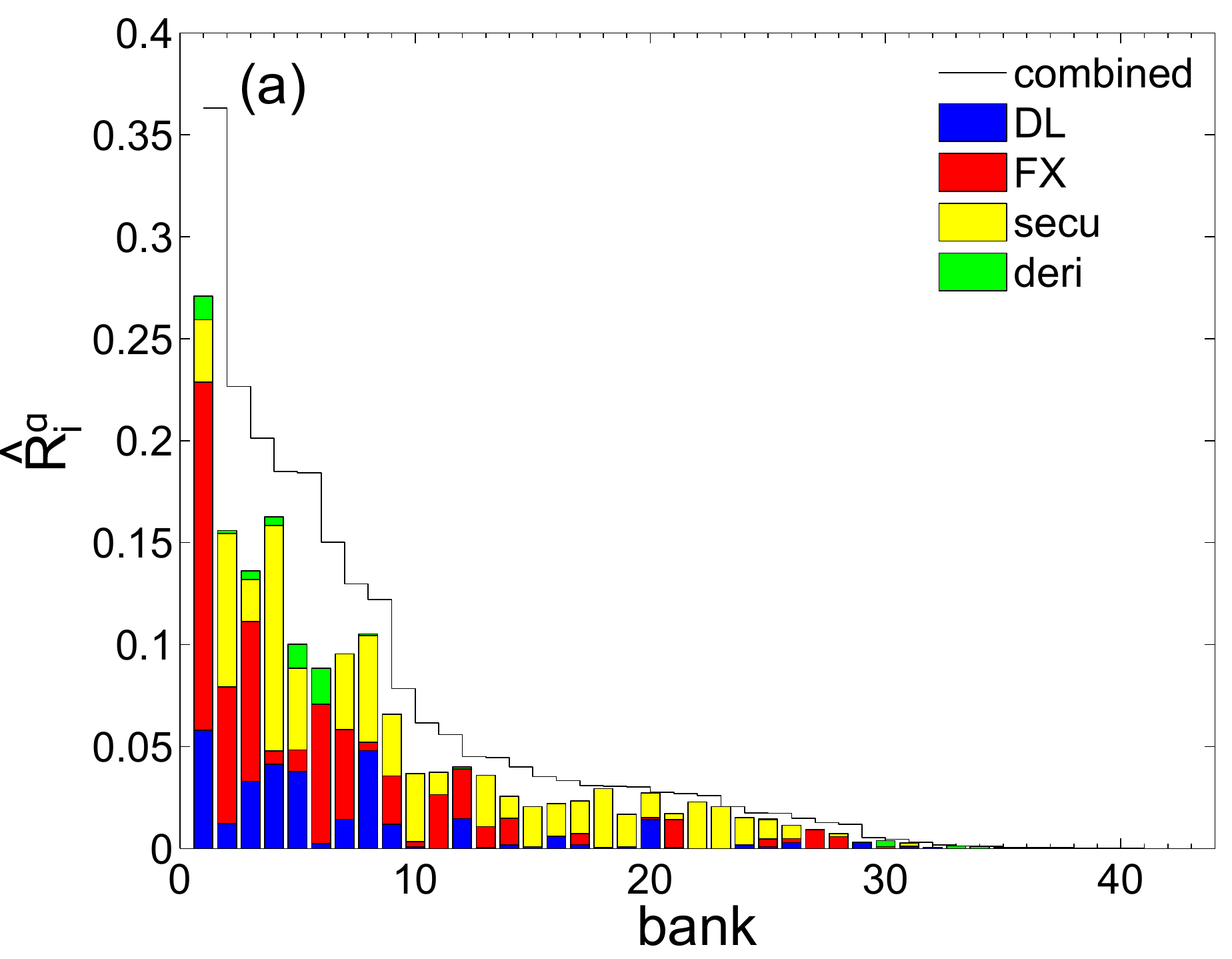} 
	\includegraphics[width=.49\textwidth]{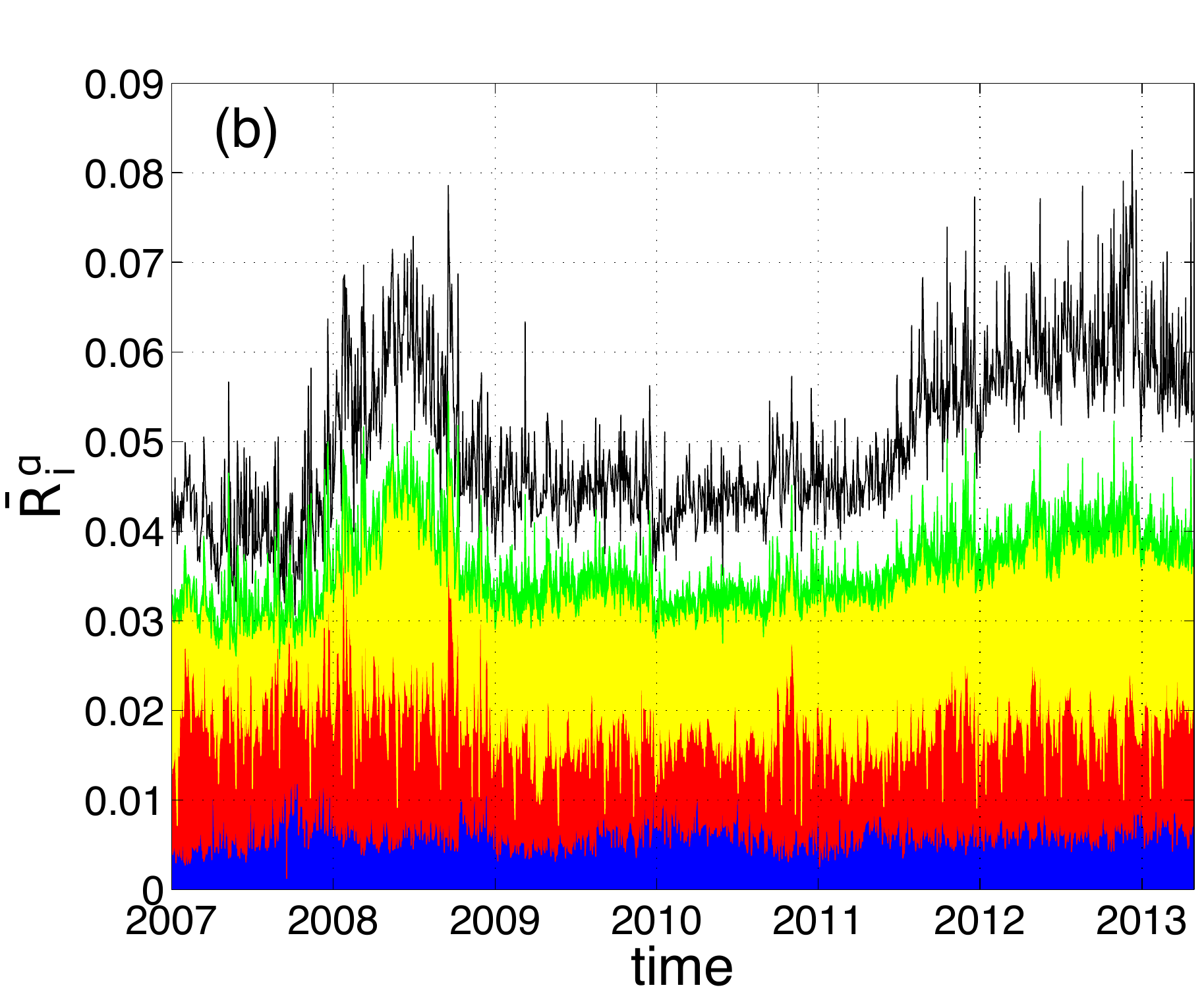} 
	\caption{(a) SR profile for the different layers. Normalized DebtRank $\hat R_i^{\alpha}$ (see \cref{multidr}) from different layers are stacked for each bank. Banks are ordered according to their DebtRank in the combined network from all layers (line). (b) Time series for the average DebtRank $\bar R^{\alpha}(t) = \frac{1}{b} \sum_{i=1}^{b} \hat R_i^{\alpha}(t)$ for all layers from Jan 2 2007 to May 30 2013. The black line shows the average DebtRank for all layers combined $\bar R^{\rm comb}(t)=\frac{1}{b}\sum_{i=1}^{b}R_i^{\rm comb}(t)$.} \label{sr_profile} 
\end{figure}

\Cref{sr_profile}(a) shows the SR-profile for the combined exposures $R_i^{\rm comb}$ (line) and stacked for different layers $\hat R_i^{\alpha}$ (colored bars) for Sept 30 2013. Clearly, individual banks have different SR contributions from the different layers, reflecting their different trading strategies. A number of smaller banks have systemic impact in the securities market only. The SR contribution from the interbank (deposits and loans) and the derivative markets is clearly smaller than the contributions from the foreign exchange and securities markets. The systemic impact of the combined layers (line) is always larger than the sum of the layers separately, $R_i^{\rm comb}>\sum_{\alpha} \hat R_i^{\alpha}$. 

\begin{figure*}
	\centering 
	\includegraphics[width=.85\textwidth]{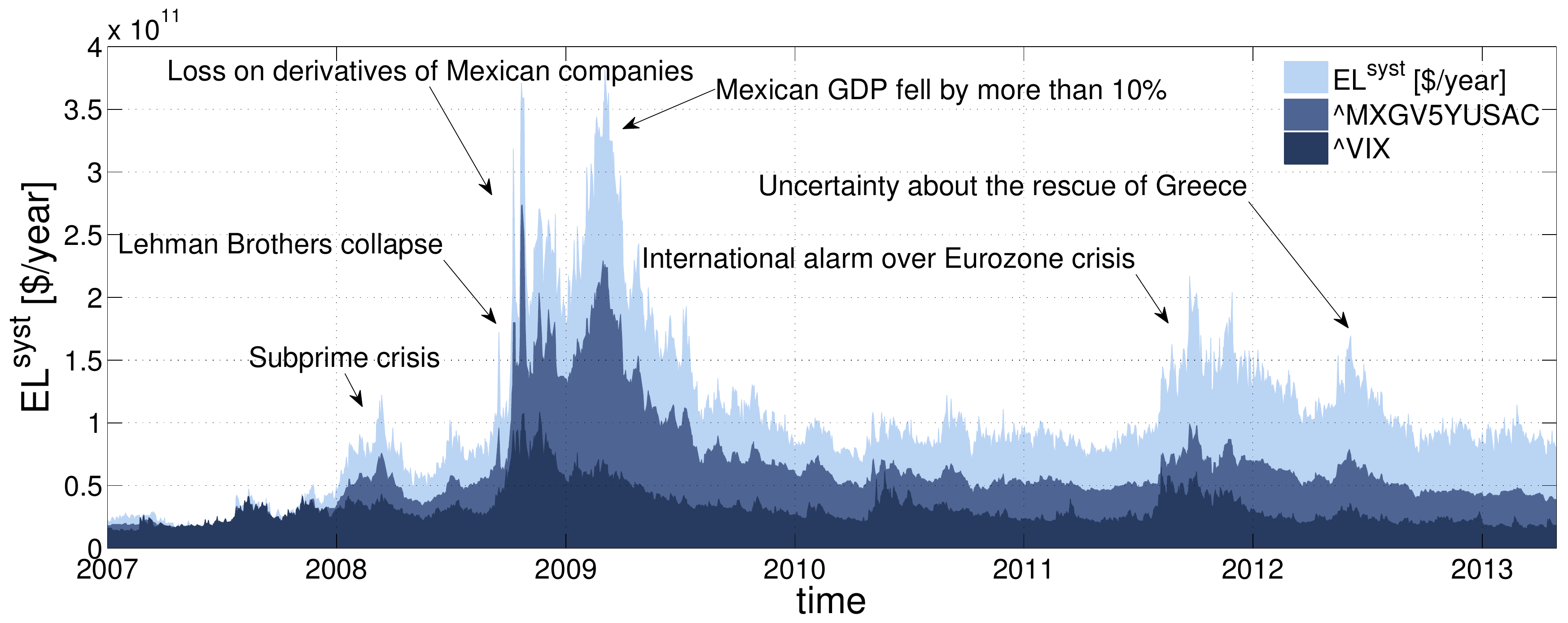} 
	\caption{Expected systemic losses ${\rm EL}^{\rm syst}$ in Mex\$ per year, in comparison to the volatility index VIX and the CDS spreads of 5-year Mexican government bonds in USD (MXGV5YUSAC). To allow comparison the MXGV5YUSAC and the VIX are scaled such that the data points coincide on January 1, 2007. Several historical events are marked. Market based indices relax fast to pre-crisis levels, whereas ${\rm EL}^{\rm syst}$ does not, indicating that the expected systemic losses are indeed driven to a large extent by network topology, and are consistently underestimated by the market. Expected losses in 2013 are about four times higher than before the crisis.} \label{timeseries_el} 
\end{figure*}

\Cref{sr_profile}(b) shows the daily {\em average DebtRank} $\bar R$ from Jan 2007-Mar 2013 for the different layers (stacked) and from the combined networks (line). As in \cref{sr_profile}(a) the combined systemic impact is always larger than the combination of all layers separately. Note that the combined average DebtRank $\bar R^{\rm comb}$ increases about 50\% from roughly 1.7 before the financial crisis of 2007-2008 to about 2.6 in 2013. The contributions of the individual exposure types are more or less constant over time. The interbank (deposits and loans) and derivative markets have smaller average DebtRank contributions than foreign exchange or securities. The derivatives market is gaining importance in Mexico after 2009. Note the relative SR increase of securities at the beginning of the subprime crisis (Dec 2007) and the subsequent decrease shortly before the collapse of Lehman Brothers. There is a marked peak in foreign exchange exposure two days after Lehman Brothers filed for chapter 11 bankruptcy protection. 

\Cref{timeseries_el} shows the daily development of ${\rm EL}^{\rm syst}$ for Mexico from 2007-2013. In Mexico there are no CDS spreads of banks available and ratings are only made available for the purpose of issuing securities by banks. This makes it difficult to derive individual default probabilities for banks. As an alternative we approximate the default probabilities for banks with sovereign default probabilities. Typically banks' strategies involve having a rating no better than the sovereign where they are registered. In general, Mexican banks are well-capitalized, especially the large subsidiaries of foreign banks. It is for this reason that we believe it is justifiable to use sovereign default probabilities as a proxy for all banks. As a reference we use the 5-year Mexican government bonds in USD (MXGV5YUSAC) and assume a 40\% recovery rate, which is the standard market convention for the quotation of CDS contracts. The short-term volatility of the expected loss is mainly driven by international events. As is the case with other credit risk models, for example models for credit default swaps \citep{Hull:2000aa,Hull:2001aa}, we assume that default events and recovery rates are mutually independent. In \cref{timeseries_el} we highlight several historical events and show the volatility index VIX and the CDS spreads scaled such that the data points coincide on January 1, 2007. We see that both the volatility index and the CDS spreads return quickly to pre-crisis levels, whereas ${\rm EL}^{\rm syst}$ clearly does not. This indicates that markets drastically underestimate SR in the system -- the expected systemic losses in 2013 are about a factor four higher than before the crisis. 

\begin{figure}
	\centering 
	\includegraphics[width=0.49\textwidth]{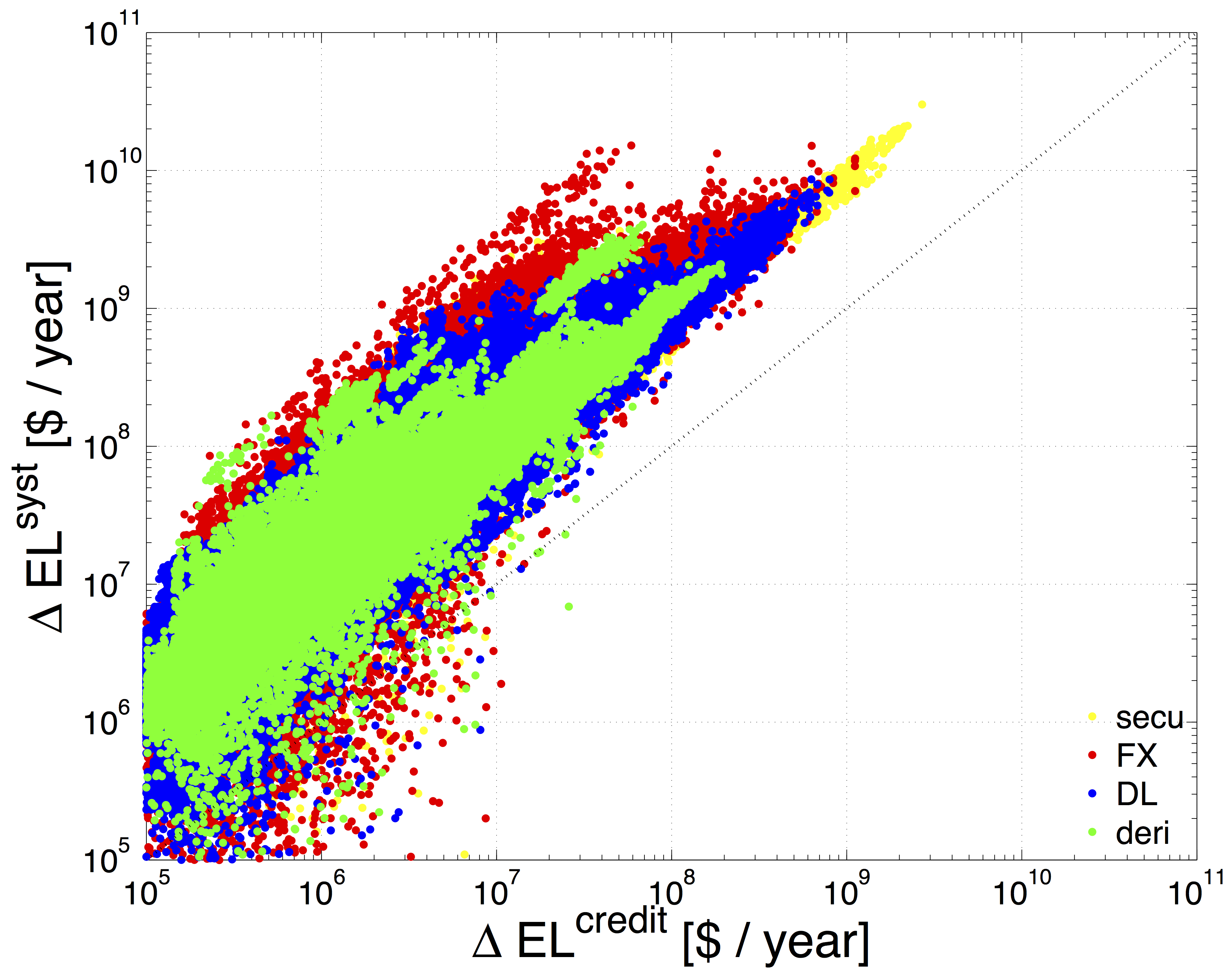} 
	\caption{Marginal increase of expected systemic loss, $\Delta {\rm EL}^{\rm syst}$, versus increase of credit risk, $\Delta {\rm EL}^{\rm credit}$, for individual exposures between institutions. Every data point represents an individual interbank liability $L_{ij}^{\alpha}$ on a given layer and given day. Data is aggregated from all banks over all days from Jan 2 2007 to May 30 2013. Exposures/liabilities lower than 10M Mex\$ are not shown. Note that $\Delta {\rm EL}^{\rm syst}>\Delta {\rm EL}^{\rm credit}$ meaning that defaults of exposures do not only affect the `lending' party but involves third parties.} \label{marginal_effects_figure} 
\end{figure}

In \cref{marginal_effects_figure} we compare the marginal contribution of individual exposures on SR and credit risk. Everyone of the ca. 500,000 individual exposures between banks across the entire time period is represented by a data point. The different layers are distinguished by colors. We immediately observe that $\Delta {\rm EL}^{\rm syst}>\Delta {\rm EL}^{\rm credit}$ for the vast majority of transactions. We made sure that this finding could not be explained by the exposure size relative to equity capital, or by capital ratios (not shown). This clearly demonstrates that marginal contributions from individual liabilities depend not only on the two involved parties, but also on the conditions of all nodes in the network. Note that small and medium-size liabilities can have SR contributions that vary by three orders of magnitude. Deposits and loans and derivatives show the lowest variability, whereas for foreign exchange it is a bit higher. Derivatives show clusters of transactions with particularly high SR contributions for the corresponding liability size. Exposures from securities cross-holdings have the highest contributions to SR. 

Note that in some cases $\Delta {\rm EL}^{\rm syst}<\Delta {\rm EL}^{\rm credit}$, meaning that a few exposures have a SR reducing effect on the network. Although counterintuitive, removing a link can change the topology of a network in a way that overall SR increases even if the total exposure of the systems is decreased. 

To exclude the possibility that this effect arises as an artifact of the measure we conducted computer simulations with the model introduced in \citet{Poledna:2014aa}. We modified (\cref{marginal_effect_simple}) to always predict an increase of SR equal or larger to the overall increase of exposure in the system, i.e. 
\begin{equation}
	\Delta^{\prime} {\rm EL}^{\rm syst} = \max\left(\Delta {\rm EL}^{\rm syst}, \Delta {\rm EL}^{\rm credit}\right) \, \quad. \label{mod_marginal_effect} 
\end{equation}
In computer simulations we used (\cref{marginal_effect_simple}) and (\cref{mod_marginal_effect}) to estimate the increase of SR in the simulated financial system. Results indicate that the unmodified version of \cref{marginal_effect_simple} predicts losses due to SR slightly better than the modified version.

\section{Conclusion} \label{discussion} 
To a large extent SR is related to the topologies of a collection of financial exposure networks (multi-layer network). This work provides,  to our knowledge, the first complete empirical picture of network-based SR in a national, system-wide context. By analyzing SR contributions from four exposure layers of the interbank network (derivatives, securities cross-holdings, foreign exchange and the interbank market of deposits and loans) we show that by relying on the single layer of deposits and loans -- as done in previous studies -- one drastically underestimates SR in the system, missing about 90\% of the total SR. We demonstrate that the exposures related to the cross-holding of securities and the exposures arising from FX transactions are crucially important components of the  SR in a country. These exposures are almost never taken into consideration as part of contagion studies, when in fact they need to be included in order to provide a more complete picture of the risks faced by the financial system.  

On a national level we suggest a SR-profile that captures the SR contributions from all financial institutions in a country across the various exposure layers. It is straightforward to use the SR profile to introduce a system-wide systemic risk measure, i.e. the average DebtRank, which takes all exposure layers into account. The average DebtRank captures the contribution of the various network topologies and can be computed on a daily scale. Interestingly, SR of the combined exposure network is higher (increasingly so over time) than the sum of SR from the four layers. This points to the non-linearity of the definition of the systemic risk measure. The root cause for this non-linearity is the propagation of shocks to financial institutions between the layers. For example, a loss from trading derivatives can spill over to other market. This non-linear effect was seen in a different context before in \mbox{\citet{Montagna:2013aa}}. If financial institutions were restricted to a particular segment of the market, e.g. some investment banks only trading securities, while other banks only engage in interbank lending, the non-linearity in the aggregation of the layers would disappear.  This evidence could provide a further argument for a structural reform of the banking system, aiming at a separation or restriction of certain trading activities as suggested in several proposals \mbox{\citep{Vickers:2011aa,Liikanen:2012aa}}. Although the idea of these proposals is to ring-fence deposit banks to safeguard against riskier banking activities, we believe that a structural separation of trading actives to separate legal entities would also reduce SR in general. 

The DebtRank, in combination with estimates of default probabilities of institutions, allows us to define a novel index, ${\rm EL}^{\rm syst}$, the {\em expected systemic losses} within a financial economy. The index quantifies the total  losses of a potential cascading event at any point in time, provided that no bail-outs are taking place at that time. This makes it possible to quantify the costs originating from SR in Pesos/Euros/Dollars per year, provided that governments would not employ a resolution mechanism (such as a bail out) for troubled banks. The expected systemic loss further enables us to compare expected costs for bailouts with the expected systemic loss, so that decisions for bailouts can be based on quantitative, transparent and rational grounds. Finally, the  expected systemic loss ${\rm EL}^{\rm syst}$ can be used to compare economies, and indicate the SR-reducing performance of policy interventions. 

We find that financial markets systematically underestimate SR. When we compare the expected systemic loss with the volatility index (VIX) and the CDS spreads of 5-year Mexican government bonds, it becomes clear that expected systemic loss follows several features of these market risk indicators. However, while the VIX returned to pre-crisis levels, and the spreads have doubled since the crisis, the expected systemic loss has quadrupled since 2007. This means that the potential direct costs for a cascading failure in Mexico would be four times higher now than before the crisis. 

In recent years various studies using multiplex network analysis have demonstrated that the understanding of a system by a single network layer can lead to a fundamentally wrong understanding of the entire system, and that the dynamics of multiplex systems can be very different from single layer networks \citep{Szell:2010aa,Nicosia:2013aa,Kim:2013aa}. The multi-layer analysis of financial networks points in a similar direction, namely that there might be much higher SR levels present in the financial system than previously anticipated, or than markets assume. There are two reasons why we still might underestimate SR. First, we do not include other potentially important sources of contagion, such as the network of overlapping portfolios and funding liquidity risk. The inclusion of more network layers is the subject of future studies. Second, in this work we assume that default events and recovery rates are mutually independent, which does not hold in practice for a number of reasons. In conclusion, true values of SR might be still significantly higher. 

\section*{Acknowledgments} The views expressed here are those of the authors and do not represent the views of De Nederlandsche Bank, Banco de M\'{e}xico or the Financial Stability Directorate. We thank B. Fuchs, C. Chrysanthakopoulos and A. Wanjek for help with the manuscript. We acknowledge financial support from EC FP7 projects CRISIS, agreement no. 288501 (65\%), LASAGNE, agreement no. 318132 (15\%) and MULTIPLEX, agreement no. 317532 (20\%).

\appendix 

\section{Classical network statistics} \label{nw_statistics_sec}
\begin{table*}
	\begin{center}
		\begin{tabular}{l l r r r r r r r}
			 & & 2007 & 2008 & 2009 & 2010 & 2011 & 2012 & 2013 \\ 
			 \multicolumn{2}{l}{Density} & & & & & & \\ \hline
			 & DL & 0.062(4) & 0.064(3) & 0.062(2) & 0.063(4) & 0.075(3) & 0.074(3) & 0.076(3) \\
			 & FX & 0.064(5) & 0.06(2) & 0.06(2) & 0.08(2) & 0.07(1) & 0.08(2) & 0.08(1) \\
			 & secu & 0.02(1) & 0.029(3) & 0.029(3) & 0.039(6) & 0.06(5) & 0.068(4) & 0.07(6) \\
			 & deri & 0.1(1) & 0.09(2) & 0.09(2) & 0.11(1) & 0.098(8) & 0.098(8) & 0.098(10) \\
			 & comb & 0.21(1) & 0.25(2) & 0.24(2) & 0.28(2) & 0.32(2) & 0.33(1) & 0.34(1) \\ \hline
			 & & & & & & & & \\
			 \multicolumn{9}{l}{Pair-wise degree correlation} \\ \hline
			 DL:Secu & All & 0.68(7) & 0.68(5) & 0.63(6) & 0.61(5) & 0.64(4) & 0.59(4) & 0.51(4) \\
			 & In & 0.46(8) & 0.49(8) & 0.43(8) & 0.42(8) & 0.49(5) & 0.46(6) & 0.38(6) \\
			 & Out & 0.53(7) & 0.58(6) & 0.49(8) & 0.46(6) & 0.49(4) & 0.36(6) & 0.31(4) \\ \hline
			 DL:FX & All & 0.54(6) & 0.47(7) & 0.39(6) & 0.4(6) & 0.35(6) & 0.36(7) & 0.38(7) \\
			 & In & 0.49(6) & 0.4(7) & 0.38(7) & 0.4(6) & 0.36(6) & 0.37(7) & 0.39(7) \\
			 & Out & 0.45(6) & 0.35(8) & 0.17(6) & 0.21(6) & 0.17(7) & 0.15(6) & 0.17(6) \\ \hline
			 DL:Deri & All & 0.32(5) & 0.24(7) & 0.2(8) & 0.13(5) & 0.25(7) & 0.29(6) & 0.27(7) \\
			 & In & 0.24(6) & 0.19(6) & 0.25(8) & 0.2(6) & 0.34(7) & 0.35(7) & 0.35(7) \\
			 & Out & 0.2(5) & 0.11(7) & 0(6) & -0.04(4) & 0.13(7) & 0.1(5) & 0.08(6) \\ \hline
			 Secu:FX & All & 0.68(7) & 0.56(6) & 0.52(8) & 0.53(7) & 0.49(5) & 0.46(5) & 0.46(6) \\
			 & In & 0.53(8) & 0.46(7) & 0.46(6) & 0.49(8) & 0.45(7) & 0.4(7) & 0.35(7) \\
			 & Out & 0.56(8) & 0.5(6) & 0.3(1) & 0.32(7) & 0.33(7) & 0.29(6) & 0.33(6) \\ \hline
			 Secu:Deri & All & 0.49(7) & 0.47(7) & 0.4(1) & 0.24(5) & 0.32(5) & 0.3(5) & 0.2(6) \\
			 & In & 0.48(7) & 0.41(9) & 0.5(1) & 0.24(5) & 0.44(8) & 0.5(5) & 0.35(9) \\
			 & Out & 0.27(7) & 0.32(7) & 0.22(8) & 0.13(6) & 0.17(8) & 0.04(5) & 0.03(6) \\ \hline
			 FX:Deri & All & 0.7(5) & 0.6(7) & 0.57(7) & 0.51(5) & 0.57(6) & 0.57(5) & 0.58(6) \\
			 & In & 0.66(6) & 0.57(8) & 0.57(7) & 0.51(6) & 0.59(7) & 0.6(7) & 0.59(7) \\
			 & Out & 0.67(6) & 0.58(6) & 0.54(7) & 0.48(6) & 0.58(6) & 0.56(6) & 0.58(7) \\ \hline
			 & & & & & & & & \\
			 \multicolumn{9}{l}{Pair-wise weight correlation} \\ \hline
			 DL:Secu & All & 0.7(2) & 0.79(9) & 0.8(1) & 0.7(1) & 0.8(1) & 0.78(8) & 0.6(1) \\
			 & In & 0.6(3) & 0.7(2) & 0.7(2) & 0.4(2) & 0.6(2) & 0.6(1) & 0.4(1) \\
			 & Out & 0.4(2) & 0.4(2) & 0.4(3) & 0.6(2) & 0.6(2) & 0.7(1) & 0.6(1) \\ \hline
			 DL:FX & All & 0.6(2) & 0.5(2) & 0.7(2) & 0.5(2) & 0.5(1) & 0.5(2) & 0.4(1) \\
			 & In & 0.5(2) & 0.5(2) & 0.6(2) & 0.5(2) & 0.5(2) & 0.5(2) & 0.4(2) \\
			 & Out & 0.5(2) & 0.4(2) & 0.4(2) & 0.3(2) & 0.3(2) & 0.3(2) & 0.3(2) \\ \hline
			 DL:Deri & All & 0.82(9) & 0.7(1) & 0.69(9) & 0.6(2) & 0.62(9) & 0.54(7) & 0.59(9) \\
			 & In & 0.6(2) & 0.6(1) & 0.5(2) & 0.5(2) & 0.7(2) & 0.5(1) & 0.4(2) \\
			 & Out & 0.6(2) & 0.4(2) & 0.4(2) & 0.3(2) & 0.2(1) & 0.2(1) & 0.2(1) \\ \hline
			 Secu:FX & All & 0.5(2) & 0.5(1) & 0.6(2) & 0.56(10) & 0.5(1) & 0.4(1) & 0.3(1) \\
			 & In & 0.5(2) & 0.5(2) & 0.7(2) & 0.7(1) & 0.6(1) & 0.5(1) & 0.5(1) \\
			 & Out & 0.3(2) & 0.2(1) & 0.2(2) & 0.09(8) & 0.09(9) & 0.03(10) & 0.03(7) \\ \hline
			 Secu:Deri & All & 0.6(2) & 0.6(1) & 0.68(10) & 0.63(8) & 0.67(7) & 0.54(3) & 0.4(1) \\
			 & In & 0.6(2) & 0.7(1) & 0.6(1) & 0.7(2) & 0.8(2) & 0.88(4) & 0.8(1) \\
			 & Out & 0.4(2) & 0.3(1) & 0.08(5) & 0.06(3) & 0.08(5) & -0.02(4) & 0.05(5) \\ \hline
			 FX:Deri & All & 0.8(1) & 0.7(1) & 0.7(1) & 0.7(1) & 0.7(1) & 0.7(1) & 0.78(8) \\
			 & In & 0.7(2) & 0.6(2) & 0.5(2) & 0.6(2) & 0.6(2) & 0.5(2) & 0.6(1) \\
			 & Out & 0.6(1) & 0.6(2) & 0.6(1) & 0.6(2) & 0.6(1) & 0.6(1) & 0.7(1)
		\end{tabular}
	\end{center}
	\caption{Classical network statistics for the multi-layer network. In the first part we show the evolution of network density for all layers combined, as well as for each individual layer. In the second part we show the pair-wise degree correlation and in the third part the pair-wise weight correlation. Values are presented as an annual average from 2007 until 2013. The number in parentheses is the standard deviation referred to the corresponding last digits of the quoted result.} \label{nw_statistics}
\end{table*}
In this section, we present classical network statistics for each of the layers, as well as for the combination of all layers. In \cref{nw_statistics}, we present the evolution of densities of different layers and the correlations between the degrees and weights of nodes across different pairs of layers. Values are presented as an annual average from 2007 until 2013.

In the first part of \cref{nw_statistics}, we show the evolution of network density for all layers combined as well as for each individual layer. Densities for the derivatives layer, deposits and loans and FX remain stable from 2007 to 2013. Densities of the securities layer and for all layers combined increase every year in the observation period.

In the second part of \cref{nw_statistics}, we show the evolution of degree correlations between layers. For every pair of layers we show the degree correlation (All), the in-degree correlation (In) and the out-degree correlation (Out). It is important to distinguish between incoming and outgoing links because they have a different economic interpretation. The former can be thought of as a form of funding or trust relationship that many banks have with one bank; the latter is related to the concept of exposure.

The third part of \cref{nw_statistics}, shows the evolution of weight correlations between layers. For every pair of layers we show the weight correlation (All), the in-weight correlation (In) and the out-weight correlation (Out).

An extensive display of stylized facts of the Mexican banking system network can be found in \citet{Martinez-Jaramillo:2014aa}. A more in-depth multiplex empirical analysis for a database similar to the one used here can be found in \citet{Molina-Borboa:2015aa}. 

\section{Multiplex network analysis} 
Given the multiplex network $L_{ij}^{\alpha}(t)$ several measures can be computed. 

\subsection{Jaccard coefficient} \label{jaccard_section} 
The Jaccard coefficient quantifies the similarity between two networks by measuring the tendency to have links present in both networks simultaneously. $J_{\alpha \beta}$ is a similarity score between two sets of elements and is defined as the size of the intersection of the sets divided by the size of their union, 
\begin{equation}
	\label{jaccard_coefficient} J_{\alpha \beta} \equiv \vert \alpha \cap \beta \vert / \vert \alpha \cup \beta \vert \quad.
\end{equation}

\subsection{Correlation coefficient} \label{corr_section} For two random variables $X$ and $Y$ with mean values $\bar X$ and $\bar Y$, and standard deviations $\sigma_X$ and $\sigma_Y$, respectively, the correlation coefficient $\rho_{X,Y}$ is defined as 
\begin{equation}
	\label{correlation_coefficient} \rho_{X,Y}=\frac{E[(X - \bar X)(Y - \bar Y)]}{\sigma_X \sigma_Y} \in [-1,1] \quad. 
\end{equation}

\section{DebtRank} \label{debtrank_section} 
DebtRank is a recursive method suggested in \citet{Battiston:2012aa} to determine the systemic relevance of nodes in financial networks. It is a number measuring the fraction of the total economic value in the network that is potentially affected by a node or a set of nodes. $L_{ij}$ denotes the interbank liability network at any given moment (loans of bank $j$ to bank $i$), and $C_{i}$ is the capital of bank $i$. If bank $i$ defaults and cannot repay its loans, bank $j$ loses the loans $L_{ij}$. If $j$ does not have enough capital available to cover the loss, $j$ also defaults. The impact of bank $i$ on bank $j$ (in case of a default of $i$) is therefore defined as 
\begin{equation}
	\label{impact} W_{ij} = \min \left[1,\frac{L_{ij}}{C_{j}} \right] \quad. 
\end{equation}
The value of the impact of bank $i$ on its neighbors is $I_{i} = \sum_{j} W_{ij} v_{j}$. The impact is measured by the {\em economic value} $v_{i}$ of bank $i$. For the economic value we use two different proxies. Given the total outstanding interbank exposures of bank $i$, $L_{i}=\sum_{j}L_{ji}$, its economic value is defined as 
\begin{equation}
	\label{ecovalue1} v_{i}=L_{i}/\sum_{j}L_{j} \quad. 
\end{equation}
Alternatively, in order to also include non interbank assets, the economic value can be defined as 
\begin{equation}
	\label{ecovalue2} v_{i}=(L_{i}+r^{loss}A^{tot}_{i})/\sum_{j}(L_{j}+r^{loss}A^{tot}_{j}) \quad, 
\end{equation}
with $A^{tot}_{i}$ as total assets excluding interbank assets of bank $i$ and a constant loss rate given default $r^{loss}=0.6$ for non interbank assets. To take into account the impact of nodes at distance two and higher, it has to be computed recursively. If the network $W_{ij}$ contains cycles the impact can exceed one. To avoid this problem an alternative was suggested in \citet{Battiston:2012aa}, where two state variables, $h_{\rm i}(t)$ and $s_{\rm i}(t)$, are assigned to each node. $h_{\rm i}$ is a continuous variable between zero and one; $s_{\rm i}$ is a discrete state variable for three possible states, undistressed, distressed, and inactive, $s_{\rm i} \in \{U, D, I\}$. The initial conditions are $h_{i}(1) = \Psi \, , \forall i \in S ;\; h_{i}(1)=0 \, , \forall i \not \in S$, and $s_{i}(1) = D \, , \forall i \in S ;\; s_{i}(1) = U \, , \forall i \not \in S$ (parameter $\Psi$ quantifies the initial level of distress: $\Psi \in [0, 1]$, with $\Psi = 1$ meaning default). The dynamics of $h_i$ is then specified by 
\begin{equation}
	h_{i}(t) = \min\left[1,h_{i}(t-1)+\sum_{j\mid s_{j}(t-1) = D} W_{ ji}h_{j}(t-1) \right] \quad. 
\end{equation}
The sum extends over these $j$, for which $s_{j}(t-1) = D$, 
\begin{equation}
	s_{i}(t) = 
	\begin{cases}
		D & \text{if } h_{i}(t) > 0; s_{i}(t-1) \neq I ,\\
		I & \text{if } s_{i}(t-1) = D , \\
		s_{i}(t-1) & \text{otherwise} \quad. 
	\end{cases}
\end{equation}
The DebtRank of the set $S$ (set of nodes in distress at time $1$), is $R^{\prime}_S = \sum_{j} h_{j}(T)v_{j} - \sum_{j} h_{j}(1)v_{j}$, and measures the distress in the system, excluding the initial distress. If $S$ is a single node, the DebtRank measures its systemic impact on the network. The DebtRank of $S$ containing only the single node $i$ is 
\begin{equation}
	\label{debtrank} R^{\prime}_{i} = \sum_{j} h_{j}(T)v_{j} - h_{i}(1)v_{i} \quad. 
\end{equation}
The DebtRank, as defined in \cref{debtrank}, excludes the loss generated directly by the default of the node itself and measures only the impact on the rest of the system through default contagion. For some purposes, however, it is useful to include the direct loss of a default of $i$ as well. The total loss caused by the set of nodes $S$ in distress at time $1$, including the initial distress is 
\begin{equation}
	\label{debtrank_self} R_S = \sum_{j} h_{j}(T)v_{j} \quad. 
\end{equation}

\section{Derivation of a practical approximation for the expected systemic loss} \label{el_approx}
\begin{figure}
	\centering 
	\includegraphics[width=.49\textwidth]{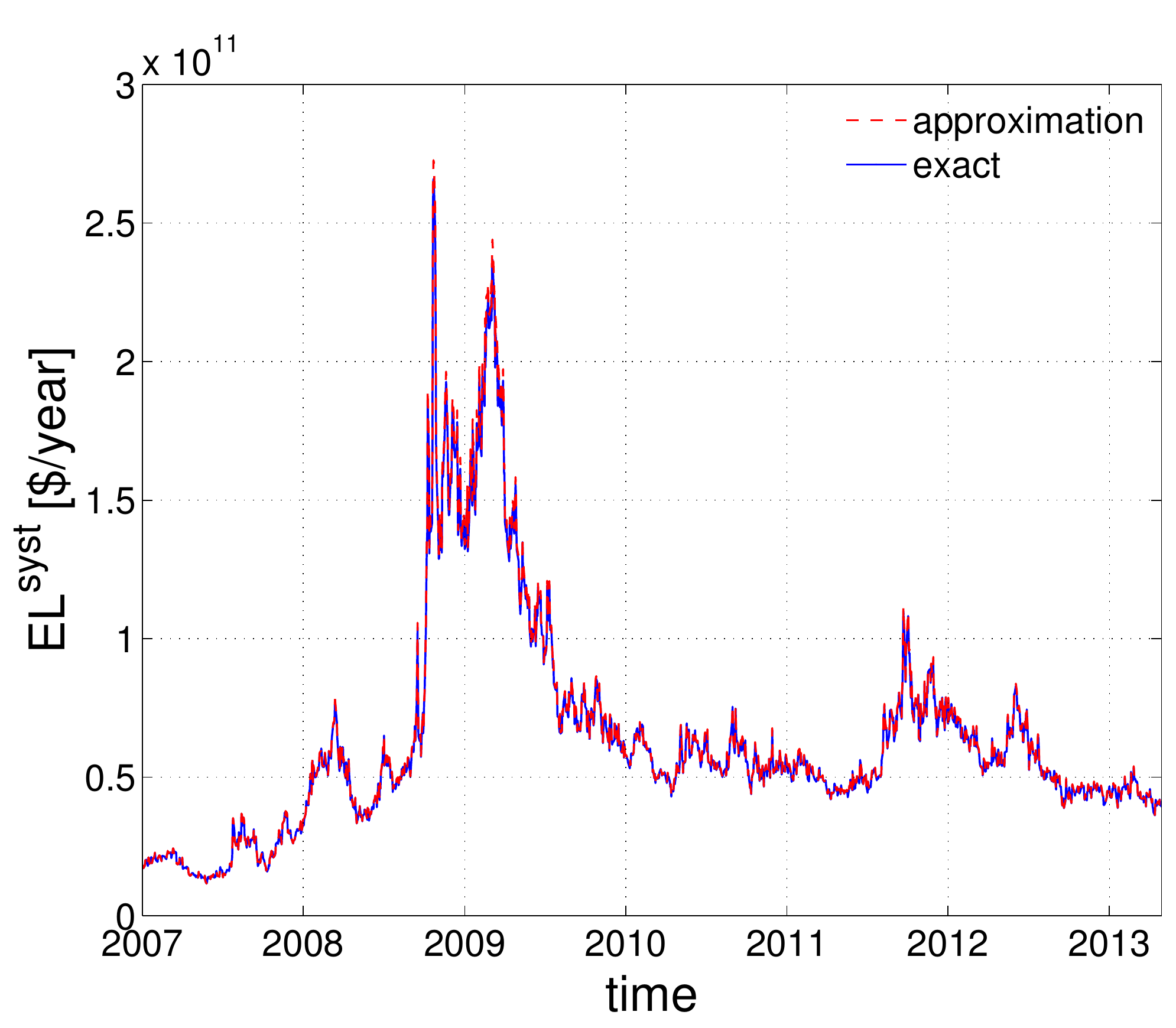} 
	\caption{Approximation for ${\rm EL}^{\rm syst}$ in comparison to the exact result given by \cref{totEL_normalized}. To allow computation in reasonable time we consider only the 15 largest Mexican banks. The approximation of \cref{totEL} is shown in red (dashed). The approximation is generally higher than the exact result (blue line). The deviation is on average less than $0.65\%$ and not more than $3.5\%$.} \label{error_fig}
\end{figure}
\Cref{totEL_normalized} is only practical for situations with less than about $20-30$ financial institutions. Computing the power set and calculating DebtRanks for all possible combinations of more than $30$ financial institutions in a large financial networks is practically unfeasible. If the default probabilities are low ($p_i \ll 1$) or the interconnectedness is low ($R_i \approx v_i$), $R_S$ can be approximated by
\begin{equation}
	R_S \approx \sum_{i \in S} R_i \quad. \label{debtrank_approx} 
\end{equation}
In an unconnected or unleveraged financial system ($R_i = v_i$), $R_S$ is exactly equal to $\sum_{i \in S} R_i$. If $p_i \ll 1$, the first terms of \cref{totEL_normalized} (with only one node initially in distress) contribute more to the final result. Thus the approximation \cref{debtrank_approx} has only a minor impact on the final result. Typically, $p_i \ll 1$ or $R_i \approx v_i$ holds in real word financial networks.
With the approximation \cref{debtrank_approx}, \cref{totEL} can be derived from \cref{totEL_normalized} by
\begin{align}
	{\rm EL}^{\rm syst} & \approx V \sum_{S \in \mathcal{P}(B)} \prod_{i \in S} p_i \prod_{j \in B \setminus S} (1-p_j) \left(\sum_{i \in S} R_i \right) \\
	& = V \sum_{i=1}^{b}\underbrace{\left(\sum_{J \in \mathcal{P}(B \setminus \{i\})} \prod_{j \in J} p_j \prod_{k \in B \setminus (J \cup \{i\})} (1-p_k) \right)}_{=1}  p_i \, R_i \label{brace} \\
	& = V \sum_{i=1}^{b} p_i \, R_i \quad.
	\label{totEL_approx} 
\end{align}
The term in brackets in \cref{brace} sums to 1 (proof by induction). \Cref{totEL_approx} is used as \cref{totEL} in the main text. This approximation is practical for large financial networks. 

We test the quality of the approximation for ${\rm EL}^{\rm syst}$ with respect to the exact result given in \cref{totEL_normalized}. In \cref{error_fig} we show the exact result for ${\rm EL}^{\rm syst}$ considering only the 15 largest Mexican banks (blue line), which can still be computed in reasonable time. The approximation of \cref{totEL} is shown in red (dashed). The approximation is generally higher than the exact result (on average less than $0.65\%$). The maximum deviation is $3.5\%$.

\end{document}